# Fabrication of nanopatterned DNA films by Langmuir–Blodgett technique

Shuxi Dai[a,b], Xingtang Zhang[a], Zuliang Du[a,*], Hongxin Dang[a,b]

[a]*State Key Laboratory of Solid Lubrication, Lanzhou Institute of Chemical Physics, Chinese Academy of Science, Lanzhou 730000, PR China*
[b]*Key Laboratory of Special Functional Materials, Henan University, Kaifeng 475001, PR China*

\* Corresponding author. Tel.: +86 378 2193262; fax: +86 378 2867282.
E-mail address: zld@henu.edu.cn (Z. Du).

**Abstract**

Fractal-like nanopatterned DNA thin films have been fabricated on mica substrate by Langmuir–Blodgett (LB) technique. Structures and components of DNA nanopatterns were investigated using atomic force microscopy (AFM) and X-ray photoelectron spectroscopy (XPS). The effect of surface pressure on the transferred DNA composite films has been studied. Scanning force microscopic observations revealed that the surface structure and morphology of DNA nanopatterns can be well controlled by changing the surface pressure. The growth mechanism of the fractal-like nanopatterns is discussed in terms of the diffusion-limited aggregation (DLA) model. The formation of large-scale DNA networks provided a well-defined template for the construction of nanocomposite films. Patterns of silver metal were prepared on DNA networks by subsequent metallization process.

## 1. Introduction

DNA is regarded as the basic building block of life. The special molecular structure of DNA has been shown to be useful in developing novel nanostructured materials [1–3]. By using DNA as building blocks or templates for the programmed assembly of nanoscale structures, the assembly of two-dimensional DNA crystals, the assembly of gold nanoclusters using DNA as the linking molecule and the formation of nanowires using DNA as template have been reported [2,3].

Recently, many research groups are attempting to fabricate DNA film, network and other patterned constructions and various methods have been used such as DNA molecular deposition, immobilization of DNA on SAMs [4,5]. However, it is hard to control the surface density and morphology of immobilized DNA. In comparison with the above methods, previous studies show that Langmuir–Blodgett (LB) technique is a very powerful tool for assembling different molecules to build ultrathin organic films with well-controlled molecular order and orientation. Many researches have focused on the interaction of DNA and surfactants at the air–water interface [6,7]. Liu et al. [8] had demonstrated the complexation of DNA and Gemini surfactants. They found different patterns can be obtained by using Gemini surfactants with different spacer length. But how to control the structures of DNA composite films is a remained problem in the application of DNA films. It is very important to study the surface morphology of the transferred DNA LB films to get more information on the film growth mechanism as well as on the nature of the complexation of DNA and surfactants.

In the present report, nanopatterned DNA LB films were fabricated by LB vertical deposition on mica at different pressures. Structures and components of DNA nanopatterns were investigated using atomic force microscopy (AFM) and X-ray photoelectron spectroscopy (XPS). The results show that large-scale fractal patterns can be fabricated through the combination of DNA and cationic surfactants by LB technique. AFM images of different DNA LB films show that the aggregated state of the DNA molecules in the composite films can be easily controlled by changing the transferred surface pressure. This method is simple but effective in controlling the fractal-like morphologies and structure of DNA composite LB films.

## 2. Experiments

### 2.1. Materials

Calf thymus deoxyribonucleic acid (DNA) and octadecylamine (ODA) were purchased from Sigma (St. Louis, MO) and used without further purification. Deionized water for all experiments was purified to a resistance of 18 MΩ cm. The DNA samples were dissolved in deionized water of 1 L, and then the aqueous solution subphase of concentration of 3 μg/ml (pH=6.5) was prepared by diluting DNA samples with deionized water. ODA was dissolved in chloroform (A.R. grade) at a concentration of 1 mM to form monolayers on the subphase.

### 2.2. Preparation of DNA LB Films and silver patterns

The preparation of the Langmuir–Blodgett films was performed using a commercial LB trough (ATEMETA LB 105, France) at room temperature under a continuous nitrogen flow. The pressure/area isotherms were recorded using a computer-controlled Langmuir film balance. The DNA/ODA composite monolayers were obtained by spreading the chloroform solutions of ODA on the DNA aqueous solutions using microsyringe. The solvent was allowed to evaporate for 20 min before the isotherms were measured. The DNA LB films were prepared according to standard vertical dipping procedures. Compressions of DNA/ODA Langmuir monolayers were performed at a speed of 1 cm/min under selected constant surface pressures. Then DNA LB films were transferred to freshly cleaved mica by vertically lifting the solid substrates with a speed of 4 mm/min through the air–water interface at a selected constant surface pressure. Then transfer the DNA LB films to a silver bath, consisting of 100 ml of silver acetate solution and 100 ml of hydroquinone solution; leave the DNA samples to soak for 2 min at room temperature. Then samples were rinsed with deionized water, and dried under nitrogen flow before AFM and XPS analysis.

### 2.3. Scanning probe microscopy (SPM)

DNA LB films and silver/DNA composite films were measured soon after preparation with a commercial Scanning Probe Microscope (SPA400, Seiko Instruments, Japan). Measurements were done in air at room temperature. A dynamic force microscope (DFM) mode with Phase Imaging was performed to avoid damage to the sample surface by the tip while scanning. DFM images were taken operating in taping mode by using microcantilevers of a spring constant of 0.2 N/m.

### 2.4. X-ray photoelectron spectroscopy (XPS)

X-ray photoelectron spectroscopy (XPS) measurements were done using an Axis-Ultra X-ray photoelectron spectrometer from Kratos (UK). Spectra were obtained at 90° takeoff angles using monochromatic Al K alpha X-rays at 150 W (15 kV, 10 mA). Pass energies of 80 and 40 eV were used for survey and high-resolution scans, respectively. Postprocessing of the high-resolution XPS spectra for peak fitting and display was done off-line using analysis Vision Processing software (Kratos, UK) running on the Sun workstation.

## 3. Results and discussion

Fig. 1 shows surface pressure–area ($\pi$–$A$) isotherms of octadecylamine (ODA) spread on pure water and aqueous DNA solutions with concentration of 3 μg/ml. The appreciable difference in the $\pi$–$A$ isotherms for ODA on different subphase can be observed. For ODA on pure water, a steep rise in surface pressure can be seen and the minimum area per molecule was obtained to be 21 Å$^2$ consisting of the molecular area of the alkyl chain. ODA on DNA subphase showing an expansion in the limiting molecular area by 45 Å$^2$, which is the evidence of the entrapment of DNA molecules from subphase and the formation of ODA/DNA mixed film. DNA is a kind of polyelectrolyte in the aqueous solution. The entrapment of DNA molecules from subphase into ODA Langmuir monolayer is driven by electrostatic interaction between the negatively charged DNA and cationic ODA molecules [6]. From previous reports, we know that the extraction from solution and formation of DNA–ODA composite films occurs without distortion to the double-helical structure of DNA [7]. So the LB technique provides a suitable method to immobilize DNA onto solid substrate and remain the native structure of DNA.

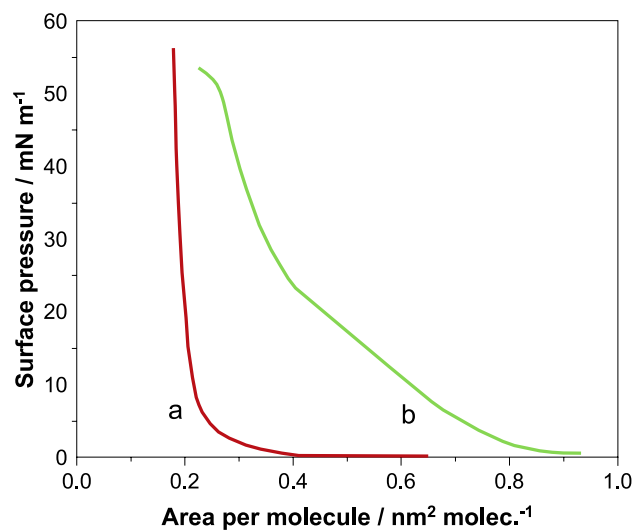

Fig. 1. $\pi$–$A$ isotherms of ODA monolayers on (a) pure water, (b) DNA=3 μg/ml in the solution at 20 °C.

Surface morphologies of ODA/DNA LB composite films transferred onto mica under different surface pressures were imaged using tapping mode DFM. Fig. 2 shows typical DFM images of monolayer ODA/DNA composite LB film transferred under surface pressure of 20 mN/m. The mica substrates are covered by two-dimensional fibers with branched network spread uniformly. We measure the height of the fibers and get an average height of 4 nm from the height profile of Fig. 2c. This value is equal to the sum height of a single layer of ODA (2 nm) and one single layer of DNA molecule (diameter of 2 nm) [9,10]. According to different AFM tips and imaging environment, the width of single calf thymus DNA strands has been obtained with values ranging from 10 to 20 nm in many previous reports [5,10]. The DNA/ODA fibers in Fig. 2 have an average width of ca. 40 nm. It indicates that the several DNA molecules aggregated under the lipid layer with their strands paralleled to the substrate surface to form a fiber seen in the AFM images. And individual DNA fiber is linking together to form such high densities of DNA network coating on mica.

To provide further information on the composition and chemical state of ODA/DNA composite LB films, we obtained the XPS spectra of the samples. A chemical analysis of ODA/DNA LB monolayer was done using XPS and the $C_{1s}$, $N_{1s}$ and $P_{2p}$ core-level spectra were recorded. The spectra obtained are shown in panels of a–c in Fig. 3. The $C_{1s}$ spectrum showed the presence of a single component at 284.6 eV and is assigned to the carbons of ODA surfactants and DNA molecules. The $N_{1s}$ spectrum showed the presence of a single component at 399.7 eV and is assigned to the nitrogens in the bases of the DNA molecules in the LB films as well as the nitrogens in the ODA molecules [11]. A $P_{2p}$ signal was recorded from the DNA LB films (Fig. 3c) and was a good indicator for the presence of DNA molecules in the composite films. The $P_{2p}$ BE agrees fairly well with reported values of DNA films immobilized on self-assembled monolayer surfaces and indicates no degradation of the DNA molecules [11,12].

In Langmuir–Blodgett technique, surface pressure is a key parameter to influence the properties of Langmuir monolayer at the air–water interface. The structure and surface density of surfactants can be well controlled just by changing the surface pressure. By changing the surface pressure, we can control the structures of DNA composite LB films easily. This method is simple but important in the preparation of DNA LB films with intricate structure and topography. In order to achieve this aim, we transferred DNA/ODA Langmuir monolayer to mica at different surface pressures. Fig. 4 presents the AFM images of DNA LB films on mica with a gradual variation of structure and surface density under surface pressures of 15, 20, 30 and 45 mN/m.

Fig. 4a shows AFM image of DNA LB film transferred under surface pressure of 15 mN/m. The mica substrates are covered by two-dimensional ODA/DNA fibers which formed branched, dendritic nanopatterns. Same patterned topography is shown in Fig. 4b, which is the AFM image of DNA LB films transferred at surface pressure of 20 mN/m. From the π–A isotherms of DNA/ODA monolayer (see Fig. 2), we can see that the mixed monolayer is in the liquid condensed state in the pressure range from 10 to 20 mN/m. The films transferred in this pressure range show a fractal dendritic growth patterns. The AFM image of DNA LB films transferred under pressure of 30 mN/m is given in Fig. 4c. From the π–A isotherms we can see that the ODA/DNA monolayer are in solid state. The branched patterns can hardly be seen. The previous DNA fibers now link together to form a closer packed network with the increase of pressure. Fig. 4d shows the AFM images of DNA LB films transferred under surface pressure of 45 mN/m. The linking network patterns of DNA fibers disappeared. Only large island-like aggregates of DNA can be seen in the overall image. Under such a high pressure, DNA fibers have twined into the coiled structures and overlapped each other to form the aggregates in the composite films.

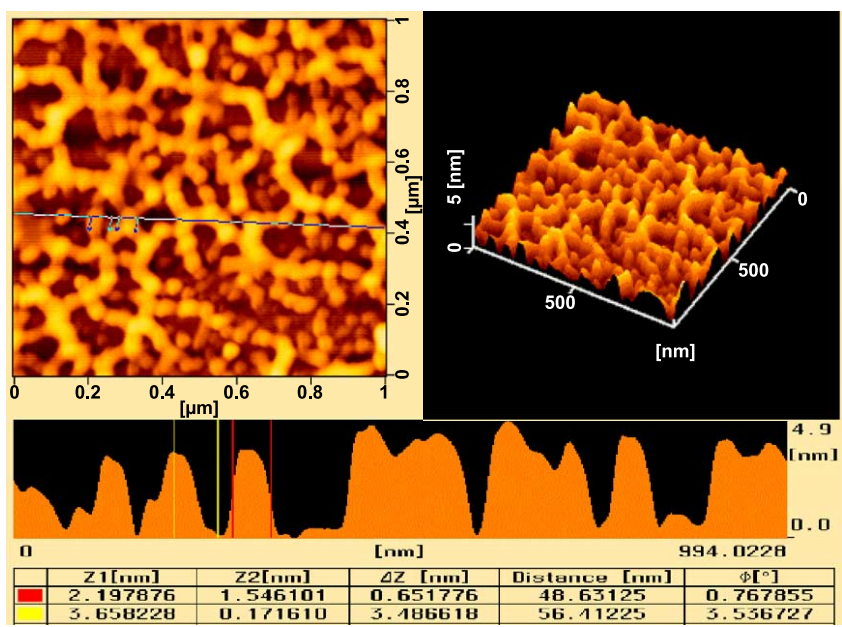

Fig. 2. AFM images of DNA/ODA LB monolayer on mica (a) AFM image, (b) 3D image, (c) height profile.

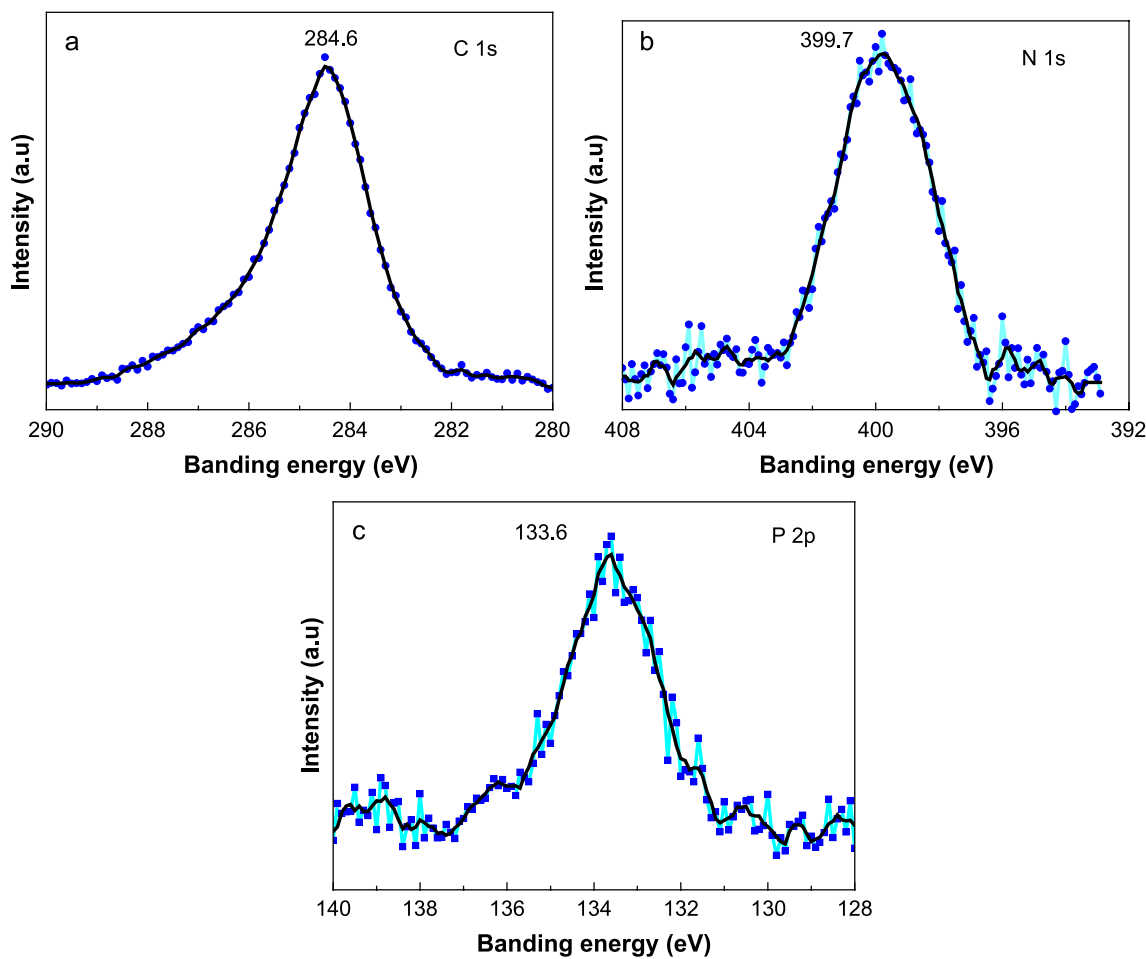

Fig. 3. XPS spectra of C 1s, N 1s, and P 2p core-level recorded from the ODA/DNA LB monolayer on mica (filled circles for raw data, thick lines for total fits).

It is clearly shown from the AFM images that the different structures of DNA patterns can be manipulated by changing the transferred surface pressures. We can also see the growth process of the DNA nanopatterns in the DNA LB films with the increasing surface pressure. It is well known that fractal structures can be modeled by diffusion-limited aggregation (DLA) and fractal structures resulting from DLA have been reported [13–15]. From the analysis in terms of DLA model, we know that the growth process of a condensed monolayer phase gives rise to dendrite formation, limited by the free diffusion of molecules in the expanded phase. Furthermore, following compression, the dendritic structures are expected to disappear through transformation into denser structures. In our experiments, the whole process of fractal formation thus includes adsorption, surface diffusion, nucleation and aggregation of DNA molecules. Initially, some adsorbed DNA molecules under ODA monolayer served as the seed nuclei of the fractal structure. The mobility of adsorbed DNA molecules then determines the formation of irreversibly fractal, nonequilibrium structures. With the increased pressure, the nucleus density of DNA is higher enough in the composite films. The possibility of generating new, large fractals will be reduced and aggregates of compact clusters will appear.

Some metal nanowires have been fabricated by using the single DNA molecule template in previous reports [3,16]. In our work, the formation of large-scale fractal DNA networks provided a well-defined template for the construction of nanocomposite films. First, patterns of silver metal were prepared on the DNA networks by subsequent metallization process. Under our experimental condition, firstly, attractive coulombic interaction between the silver ions and the negatively charged phosphate backbone of DNA molecules leads to the attachment of silver ions to the DNA LB films. Then silver nanoparticles were formed just along the DNA fibers with the reduction of the silver ions by hydroquinone solution. Using this method, many metal and semiconductor composite nanopatterns can be fabricated by using the template of DNA LB films.

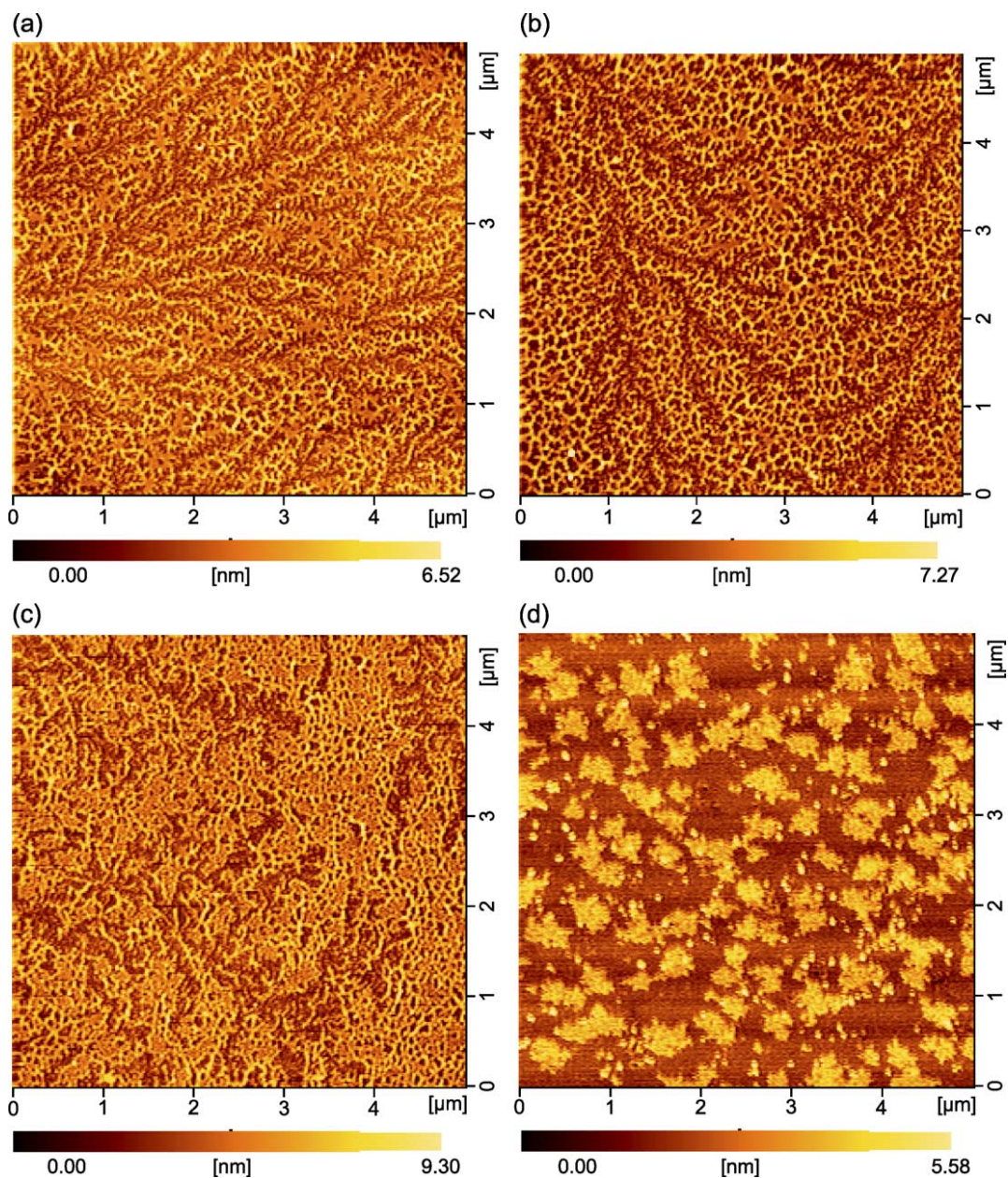

Fig. 4. AFM images of DNA/ODA LB monolayer transferred on mica substrate under different surface pressures (a) 15 mN/m, (b) 20 mN/m, (c) 30 mN/m, (d) 45 mN/m.

Fig. 5 shows DFM images (topography and phase) of thin silver film chemical deposited on DNA LB film transferred under 20 mN/m. The topography image shows that the surface was covered by two-dimensional metal fibers with uniform network structures. The images show that silver-coated fibers intertwined together to form a silver metal patterning. From the cross sections analysis of Fig. 5c we obtain an average width of 80 nm and height of 10 nm for silver-coated metal fibers, which are much larger than those data obtained form DNA fibers in Fig. 2c. From comparison of the AFM images of DNA LB films and silver metal films, we can see that the net heights of silver coating along DNA molecules is about 5–7 nm. It can be seen clearly that silver metal deposition occurs on the patterned DNA molecules. XPS spectra for P 2p and Ag 3d for silver metal films are showed in Fig. 6. The positions and intensity for P 2p and Ag 3d were compared for DNA LB films and silver metal patterning. It is clearly show that the Ag 3d peaks were present and the P 2p peaks were absent. The Ag $3d_{3/2}$ and Ag $3d_{5/2}$ peaks are identified at 374.2 and 368.2 eV, respectively. The XPS signal established that the particles patterned on DNA LB film were primarily silver metal ($Ag^0$). It indicated that the silver patterns grew only along the DNA fibers and remained the fractal network structure of DNA LB films.

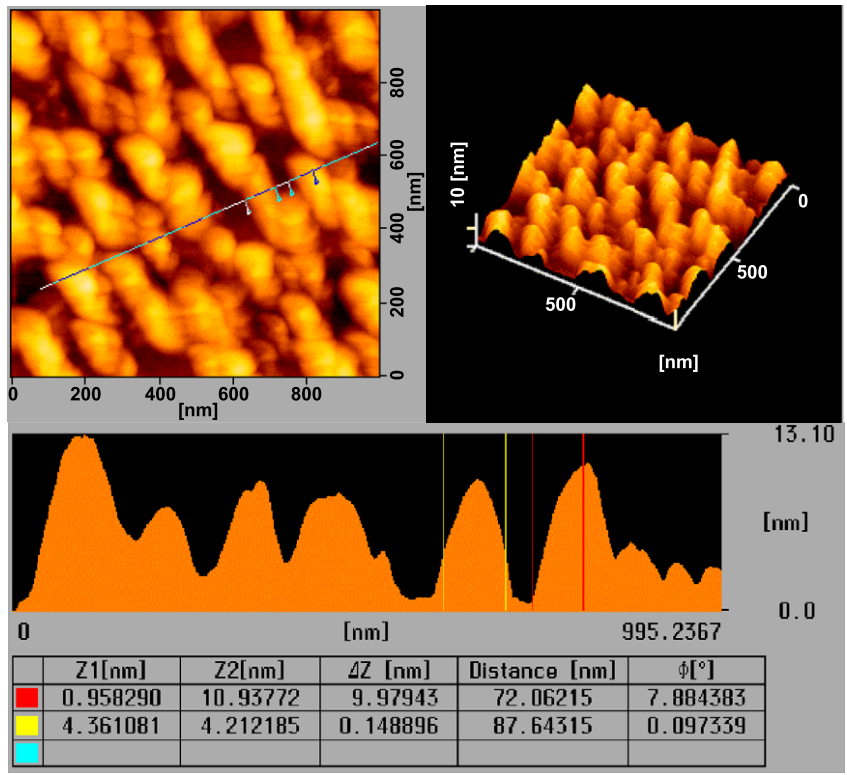

Fig. 5. Topography image of silver metal pattern chemical deposited on DNA LB film transferred under 20 mN/m of 1000×1000 nm² (a) height image, (b) 3-D image, (c) line profile.

## 4. Conclusion

The present work provides a controlled and precise way to fabricate DNA-based molecular devices. Structures and components of DNA nanopatterns were monitored using AFM and XPS. It shows that we can construct DNA fractal nanopatterns with precise control of its structure and properties using LB technique. Patterns of silver metal were prepared using DNA LB films templates by subsequent metallization process. It is shown that the use of DNA templates can construct large-scale nanocomposite films, which are promising material in the applications of nanoscale electronics. Further work will focus on the investigation of using such DNA patterns as 2-D template or scaffold for the construction of novel structured DNA-based nanomaterials.

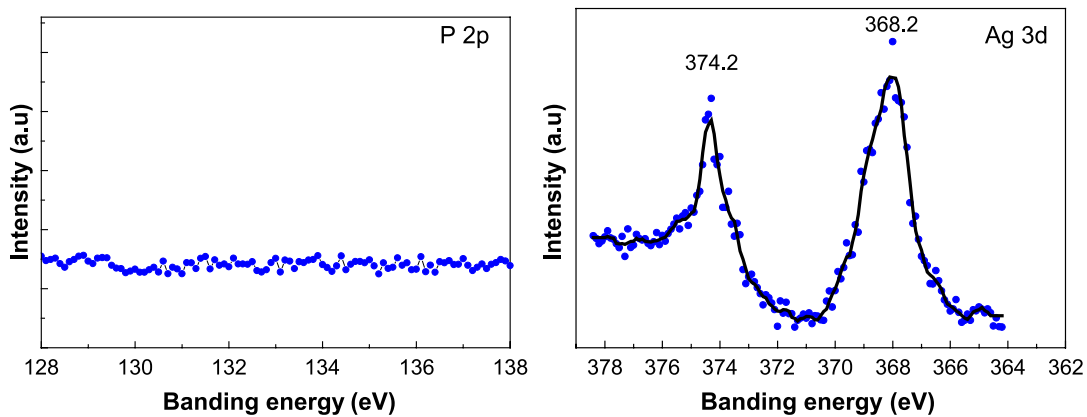

Fig. 6. XPS spectra of P 2p and Ag 3d core-level recorded from the silver metal patterns based on DNA LB film templates (filled circles for raw data, thick lines for total fits).

## Acknowledgements

This work was supported by the Natural Science Foundation of China (No. 90306010 and 20371015) and the State Key Basic Research "973" Plan of China (No. 2002CCC02700).